\documentclass[a4paper,fleqn]{cas-sc}

\usepackage{amsmath, amssymb, amsfonts, dcolumn, graphicx, float, hyperref, xcolor, placeins, multirow, hhline, threeparttable}
\usepackage[authoryear]{natbib} %longnamesfirst

\def\tsc#1{\csdef{#1}{\textsc{\lowercase{#1}}\xspace}}
\tsc{WGM}
\tsc{QE}

\defcitealias{ks91}{KS91}
\defcitealias{young:1994}{Y94}
\begin{document}
\let\WriteBookmarks\relax
\def\floatpagepagefraction{1}
\def\textpagefraction{.001}

\shorttitle{Solar image quality assessment}
\shortauthors{C.W. So, et. al.}
\title[mode = title]{Solar image quality assessment: a proof of concept using Variance of Laplacian method and its application to optical atmospheric condition monitoring}

\author[1]{Chu Wing So}
\ead{cwso@lcsd.gov.hk}
\cormark[1]

\author[2]{Edwin Lok Hei Yuen}
\ead{lhyuene@connect.hku.hk}

\author[1]{Edgar Heung Fat Leung}
\ead{edgarleung0716@gmail.com}

\author[2]{Jason Chun Shing Pun}
\ead{jcspun@hku.hk}

\affiliation[1]{organization={Hong Kong Space Museum},
            addressline={10 Salisbury Road, Tsim Sha Tsui, Kowloon}, 
            city={Hong Kong},
            country={China}}

\affiliation[2]{organization={Department of Physics, The Chinese University of Hong Kong},
            addressline={Sha Tin, New Territories}, 
            city={Hong Kong},
            country={China}}           

\begin{abstract}
Here we present a proof of concept for the application of the Variance of Laplacian (VL) method in quantifying the sharpness of optical solar images.
We conducted a comprehensive study using over 65,000 individual solar images acquired on more than 160 days. 
Each image underwent processing using a VL image processing algorithm, which assigns a `score' based on the sharpness of the solar disk's edges. We studied the scores obtained from images acquired at different conditions. 
Our findings demonstrate that the sharpness of the images exhibits daily trends that are closely linked to the altitude of the Sun at the observation site. We observed a significant degradation in image quality only below a certain altitude threshold. Furthermore, we compared airmass formulae from the literature with our sharpness observations and concluded that the degradation could be modeled as an Image Sharpness Function (ISF), which exhibits similarities to airmass variations.
In addition to assessing image quality, our method has the potential to evaluate the optical atmospheric conditions during daytime observations. Moreover, this technique can be easily and cost-effectively applied to archival or real-time images of other celestial bodies, such as the Moon, bright planets and defocused stars.
Given that ISF is unique to each location and sensitive to sky conditions, the development of an ISF is not only beneficial for routine observation preparation but also essential for long-term site monitoring.
\end{abstract}


\maketitle

\section{Introduction} \label{sec:intro}
The quality of astronomical data has garnered significant attention due to its impact on scientific discoveries. Higher-quality images with improved spatial resolution have the ability to reveal finer details within celestial objects. This has been exemplified in the resolved images of notable red giants such as Betelgeuse \citep{burns:1997} and W Hydrae \citep{vlemmings:2021}. Additionally, the study of exoplanets has benefited from higher-quality images, as demonstrated in the case of 2M1207b \citep{patience:2010,stone:2012,ricci:2017}.

In ground-based optical observations conducted under turbulent atmospheric conditions, the quality of the acquired images can be significantly affected by the attenuation of starlight due to scattering and absorption \citep{coulman:1985}. To mitigate the impact of atmospheric extinction, astronomers often choose to observe their targets when the zenith angle is small, corresponding to a lower airmass. For the same reason, observatories are typically established in locations with dry or thin-air conditions to minimize atmospheric interference \citep[e.g.][]{giovanelli:2001,yoshii:2016}. In parallel, remarkable progress has been made in developing techniques to overcome the attenuation of starlight caused by the atmosphere. One advancement is the implementation of adaptive optics \citep{davies:2012} which enables real-time corrections for atmospheric distortions, resulting in improved image quality and higher spatial resolution.

Astronomical seeing is a widely used metric for characterizing atmospheric conditions during observations. It is commonly quantified by measuring the stellar image's Full Width at Half Maximum (FWHM), which represents the angular size of the point spread function of a stellar image on the sensor \citep{burns:2022}. A smaller FWHM corresponds to better seeing conditions, indicating a smaller atmospheric turbulence. Conversely, a larger FWHM indicates poorer seeing conditions with increased atmospheric distortion. In the visual band, the typical size of the seeing disk at the world's best observational sites during photometric nights is approximately 0.6" to 0.8" \citep{skidmore:2009,ramio:2012,ma:2020}. These sites are renowned for their exceptional atmospheric conditions, enabling high-quality observations. In contrast, suburban sites located near coastal metropolitan cities typically experience significantly larger seeing disks, typically around 10 times bigger \citep{so:2010}. The larger seeing disks at such locations are attributed to the increased atmospheric turbulence caused by local factors like urbanization and proximity to the coast. 

The traditional seeing measurement from starlight is not applicable to daytime observations, such as solar observations. Attempting to represent diurnal conditions using measurements derived from nocturnal seeing is far from optimal, primarily due to the dynamic and rapidly changing nature of atmospheric optical properties caused by various natural and artificial factors \citep{perrone:2022}. While there are sophisticated solutions and specialized devices available for daytime seeing measurements, they often require substantial financial, temporal and labor resources. For instance, the solar equivalent of a Differential Image Motion Monitor (DIMM) involves the use of a specially designed telescope with two apertures and a prism to split the light from the solar limb. The seeing condition is then quantified by measuring the differential motion between images of the solar limb \citep[e.g.][]{beckers:2001,ozisik:2004,kawate:2011,irbah:2016,song:2018}. Once the setup is installed, the telescope becomes dedicated solely to seeing monitoring. This necessitates the investment in an additional set of equipment, including the telescope, mount, detector and other necessary components. Given the significant investment of resources required by these dedicated devices, their widespread adoption for routine daytime seeing measurements is not always feasible. As a result, alternative approaches or methodologies that are more accessible and cost-effective are sought for daytime seeing assessment in solar observations.

It is widely recognized that good atmospheric seeing conditions result in sharper images, while poor seeing conditions lead to image blurring. Building upon this relationship, we aim to assess image quality and variations in daytime seeing by directly analyzing the sharpness or blurriness of solar images. In the present study, we investigate the potential of applying a sharpness detection algorithm to solar images for this purpose. Through preliminary tests conducted on a dataset of over 65,000 individual solar images, we demonstrate the practicality of our approach.
We observe a correlation between solar altitude and image sharpness, where higher solar altitudes generally correspond to sharper images and vice versa. This trend can be modeled in a manner similar to airmass variation, providing insights into the variations in daytime seeing.
Importantly, our method does not require any additional resources beyond the existing solar image archive. It leverages the available data and makes efficient use of the resources already in place. Moreover, the potential of our method extends beyond solar images and can be applied to other celestial objects such as the Moon, bright planets and defocused stars. 

This paper is structured as follows: Sect.~\ref{sec:methods} provides an introduction to the sources of the solar images used in this study and outlines the methodology of the sharpness detection algorithm. Subsequently, in Sect.~\ref{sec:results_analysis}, we present the data quality control, statistics and results obtained from applying the algorithm to the solar images and provide a detailed analysis of the findings. Finally, in Sect.~\ref{sec:discussion}, we draw conclusions based on the results and engage in a comprehensive discussion of the implications and potential applications of our method in the assessment of image quality and variations in daytime seeing.

\section{Methods}\label{sec:methods}
\subsection{Solar Telescope and Images}\label{sec:telescope_image}

\begin{figure}
	\includegraphics[width=\columnwidth]{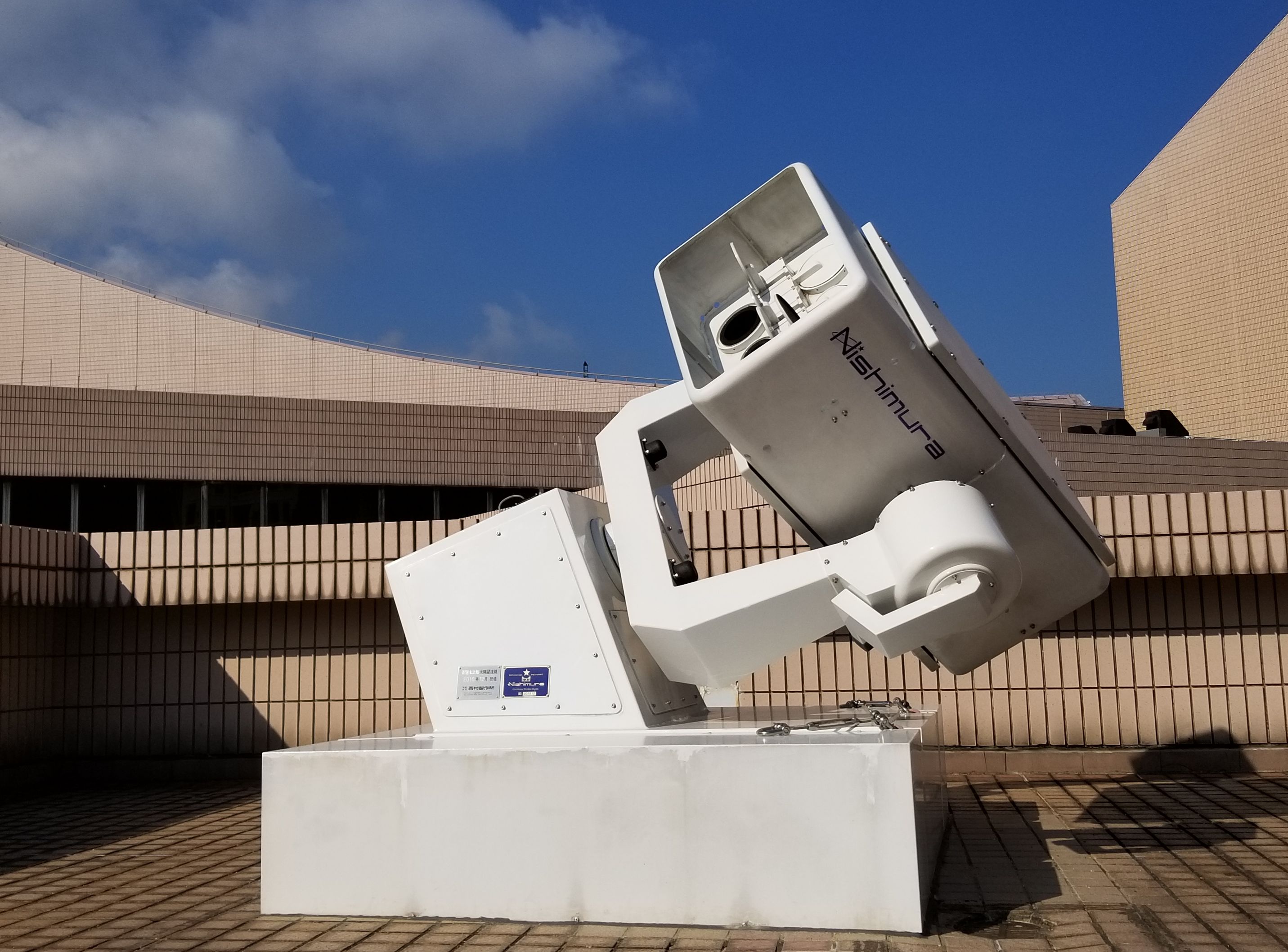}
    \caption{
    This photograph shows the Hong Kong Space Museum's Weatherproof Vacuum Solar Telescope System (STS) and provides a glimpse of its surroundings. The Hong Kong Cultural Centre, located to the west of STS, is partially visible in the background. As discussed in Sect.~\ref{sec:statistics}, the presence of the cultural center has imposed certain limitations on the unobstructed view of the Sun.}
    \label{fig:STS}
\end{figure}

\begin{figure}
	\includegraphics[width=\columnwidth]{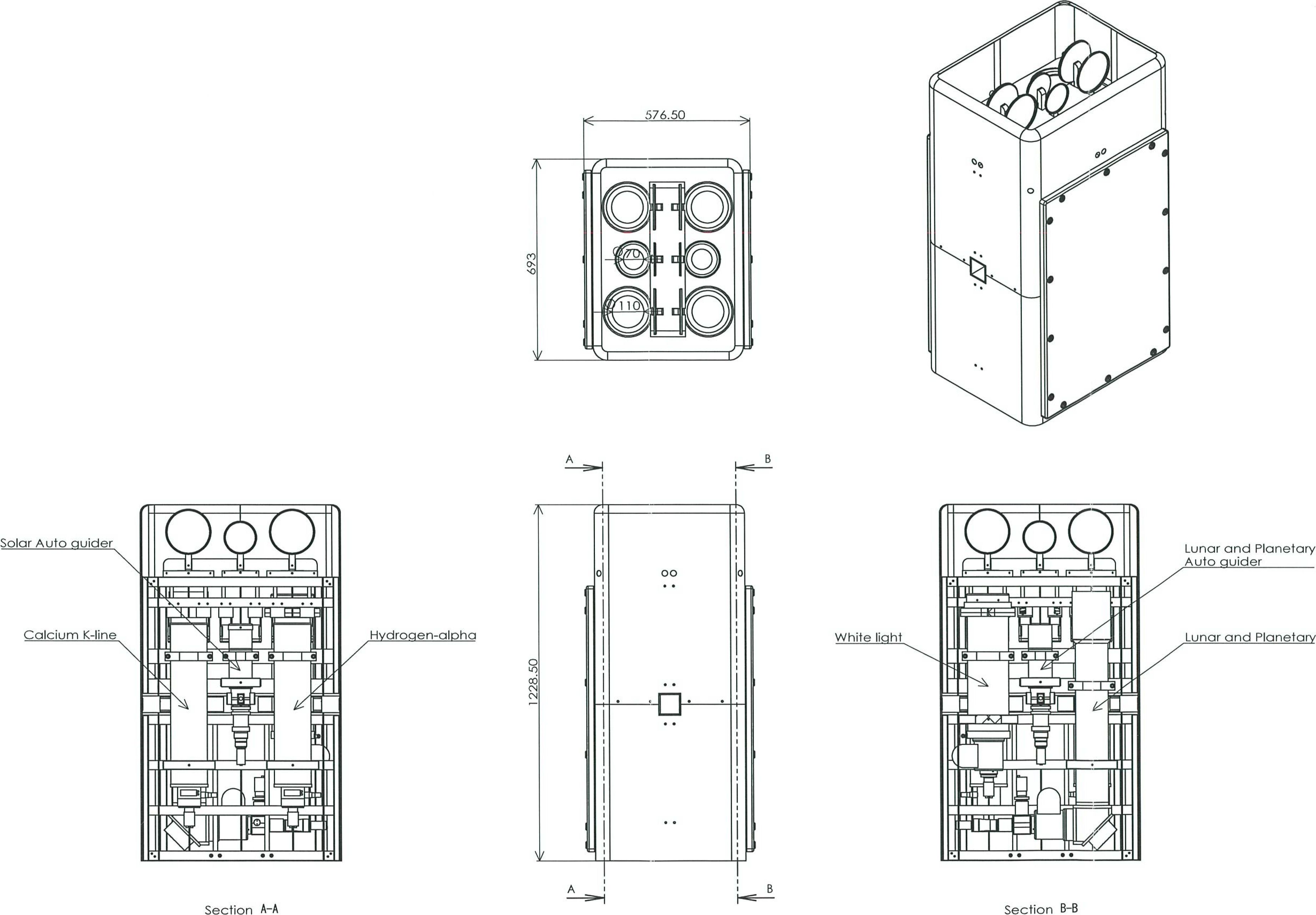}
    \caption{Technical drawing of STS}
    \label{fig:technical}
\end{figure}

We obtained pseudo white light solar images (with a peak at 430 nm) using the Weatherproof Vacuum Solar Telescope System (STS) at the Hong Kong Space Museum. STS is manufactured by Nishimura Co,. Ltd. in Japan. Permanently installed on the museum's roof (located at 22.29407$^\circ$N, 114.17146$^\circ$E) in the urban harbor front of Hong Kong, STS provides a reliable platform for conducting solar observations. Refer to Fig.~\ref{fig:STS} for a photograph of STS. In addition to white light observations, STS has been utilized since 2017 to observe the Sun in Hydrogen-alpha (with a peak at 656.3 nm) and Calcium-K (with a peak at 393.3 nm) bands using separate telescopes. Furthermore, STS includes a lunar cum planetary telescope and two auto-guiders. The six refracting telescopes in a 2$\times$3 array are mounted on an equatorial fork mount, as depicted in the technical drawing shown in Fig.~\ref{fig:technical}.

Under favorable weather conditions, the auto-guider effectively maintains precise tracking of the Sun by automatically tracing the solar disk's edge. The presence of occasional cloud passages within the field of view does not significantly impact the quality of guiding. However, during prolonged periods of cloudy weather, STS may temporarily lose track of the Sun, necessitating manual re-centering.

Each sun-observing telescope of STS has an aperture of 100 mm and a focal ratio of F/5. Each telescope is fitted with a color CCD camera that captures an image of the solar disk every minute (every 5 minutes since 2023 January 19). These images are saved in JPG format for archival purposes. The image has a resolution of 2,448$\times$2,050 and a dynamic range of 8 bits. Each pixel sizes 4 $\mu$m$\times$4 $\mu$m. The effective field of view of the sensor is enough to cover the entire solar disk throughout the year. In addition to archival purposes, the images are live-streamed to the museum’s gallery and the public website\footnote{\url{https://hk.space.museum/SolarTelescope/SolarTelescope.html}}.

\subsection{Variance of Laplacian}\label{sec:VL}

The Variance of Laplacian (VL) method is a blur detection algorithm used in computer vision. It was originally developed for digital image analysis \citep{adrian:2015}. The method involves convolving the image with a 3$\times$3 Laplacian operator and calculating the variance, which is the squared standard deviation. The Laplacian operator is commonly used for edge detection by measuring the second derivative of the image's intensity and highlighting regions with rapid intensity changes \citep{wang:2007}. Blurred images tend to have fewer edges, resulting in a lower variance. On the other hand, sharper or well-focused images have a higher variance, indicating a wider spread of edge-like and non-edge-like responses. The VL method is effective in determining the level of blurriness in an image based on its variance. The Laplacian operator has also been applied as an auto-focusing technique in microscopy due to its reliability and speed \citep{pech:2000}.

Solar images, especially white light images, are ideal for edge detection. It is because solar images often contain a featureless symmetric disk with a clearly defined and high-contrast edge. Sunspots, which are captured in some solar images, also create edges that can affect the VL of the image. However, sunspots generally move slowly relative to the integration time of the image and their impact on the VL is similar to the factors affecting the solar disk. As a result, there is typically no systematic difference in the VL of images with and without sunspots.

\begin{figure}
	\includegraphics[width=\columnwidth]{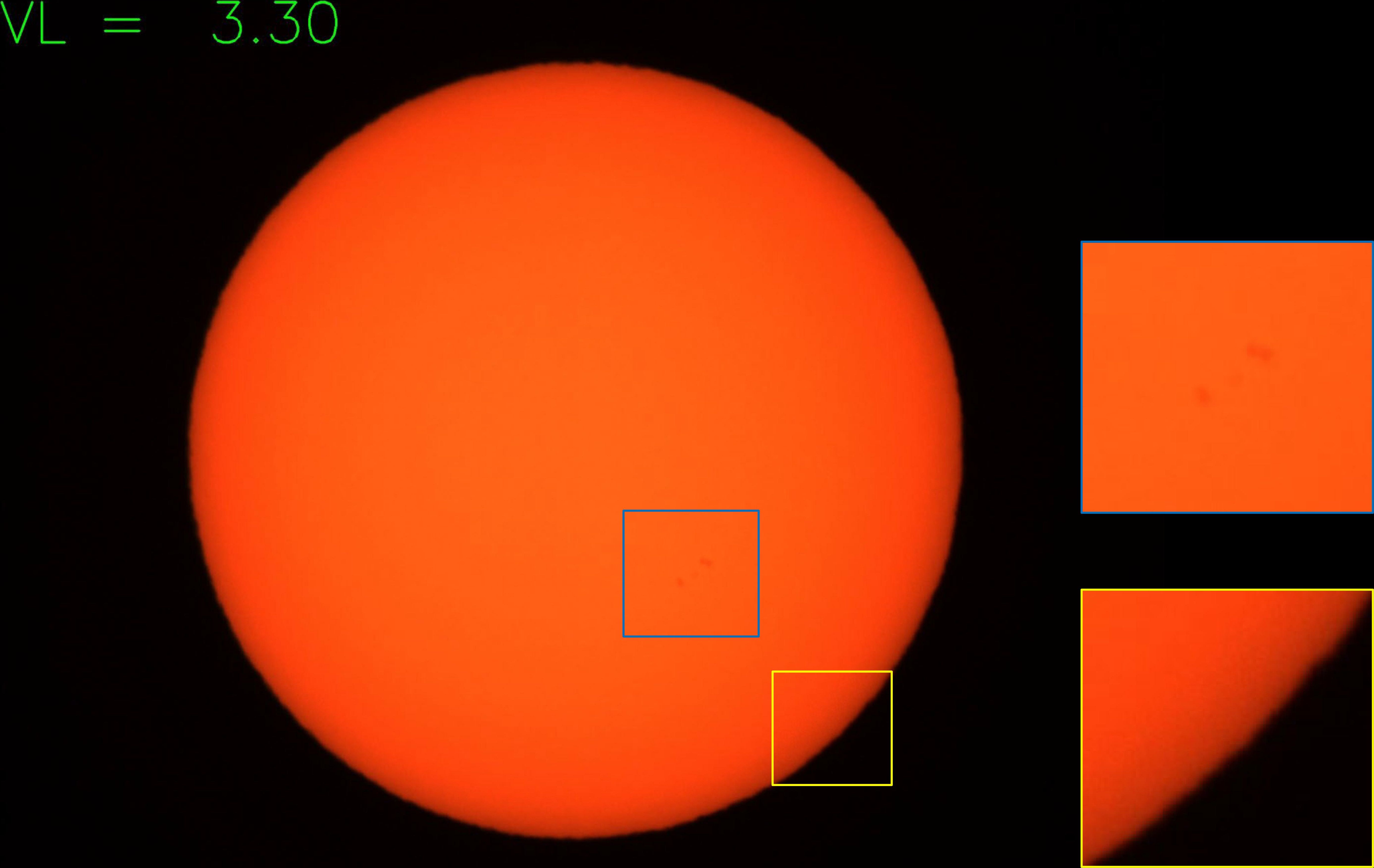}
    \caption{Here is an example of a blurry solar image with a smaller VL value. The zoomed-in sub-images highlight the sunspots and the edges of the solar discs in detail.}
    \label{fig:sample_images_blur}
\end{figure}

\begin{figure}
	\includegraphics[width=\columnwidth]{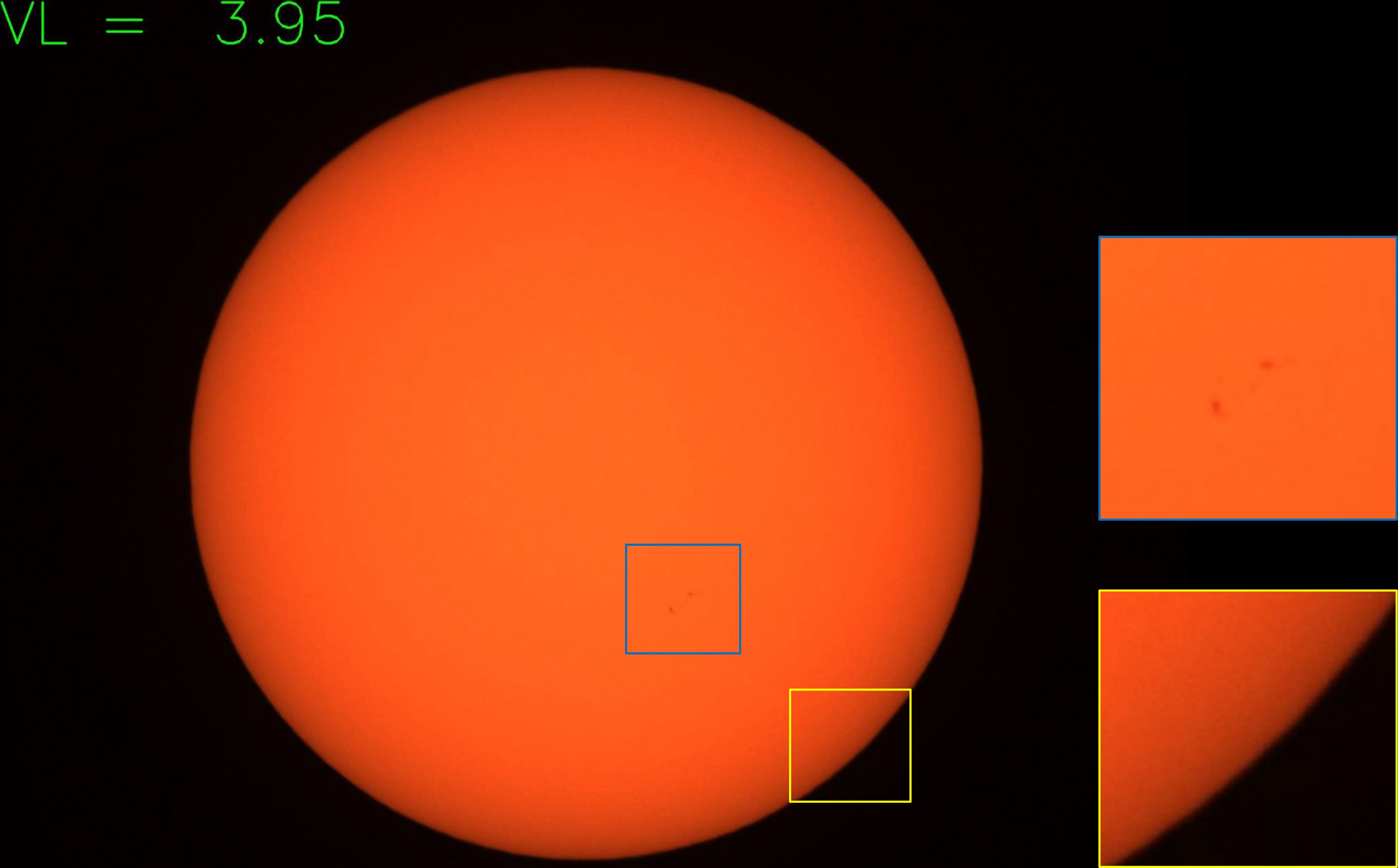}
    \caption{Same as Fig.~\ref{fig:sample_images_blur} for a sharp image with a larger VL value.}
    \label{fig:sample_images_sharp}
\end{figure}

To compute the VL from batches of solar images, we developed a \textsc{Python} program. We provide a pair of sample images in Fig.~\ref{fig:sample_images_blur} and~\ref{fig:sample_images_sharp}. In Fig.~\ref{fig:sample_images_blur}, the solar disk's edge and the sunspots appear blurrier, indicating a smaller VL value. On the other hand, in Fig.~\ref{fig:sample_images_sharp}, the same features appear sharper, indicating a larger VL value. This demonstrates the relationship between the VL value and the clarity of solar images.

\section{Results \& Analysis}\label{sec:results_analysis}
\subsection{Data quality control \& statistics}\label{sec:statistics}

A total of 288,756 individual images captured between 2021 January and 2023 June were downloaded from the archive. To ensure data quality, a series of selection criteria were applied in the following sequence, taking into account the system operation and the Sun's position:

\begin{enumerate}
    \item Exposure time: Images were selected if they were exposed after sunrise and before sunset. This criterion was implemented to exclude images captured during periods when the operator forgot to shut down the system after sunset.
    \item Unobstructed Sun: Images were filtered to ensure that the Sun was not blocked by nearby buildings during exposure. A site survey was conducted and it was determined that the telescopic view to the Sun's path below 30$^\circ$ altitude in the azimuthal direction between 205$^\circ$ and 300$^\circ$ may be obstructed by the Hong Kong Cultural Centre (see Fig.~\ref{fig:STS}).
    \item Exclusion of initial images: The first few images captured within the first 5 minutes of each day were excluded. These initial images were used for testing purposes, during which the operator performed tasks such as image centering and focusing. Manual checks on sample images revealed that these tasks were typically completed within 5 minutes. 
\end{enumerate}

The previous selection criteria resulted in the exclusion of approximately 1.5 per cent of the archival data. The remaining images were exposed under varying cloud conditions, ranging from overcast to clear. Some images appeared completely black due to overcast conditions, while others had cloud patches covering only a portion of the Sun. On the other hand, a subset of the images showed a clear and distinct view of the Sun. To minimize the influence of black and cloudy images effectively and to expedite the selection process without relying solely on subjective and time-consuming manual tasks, two additional objective selection criteria were implemented: 
\begin{enumerate}
    \setcounter{enumi}{3}
    \item Black pixel threshold: Images were selected if the total number of black pixels was smaller than 80 per cent. The statistic was calculated using \textsc{Python} codes. By setting a threshold on the amount of black pixels, it aimed to exclude images that were excessively dark or had a significant portion of the image obscured.
    \item Daily sunshine percentage: Images were further refined based on the sunshine duration. The daily sunshine duration data, measured by the Hong Kong Observatory \citep{hko:2021,hko:2022,hko:2023}, were used to calculate the percentage of time during the day that the Sun was visible. Images were selected if the daily sunshine percentage exceeded 80 per cent. This criterion aimed to ensure that the selected images were captured during periods of favorable weather conditions with a significant amount of sunshine, providing a sizable number of images per day for the calculation of DVL described in Sect.~\ref{sec:DVL}.
\end{enumerate}

It is understandable that setting stricter selection criteria may result in a smaller number of available images for analysis. Conversely, including cloudy images could potentially impact the accuracy of sharpness detection. The determination of the cut-off values for these criteria is typically based on tests conducted on real images, striking a balance between data quality and sample size. In the end, 80.8 per cent of images passed criterion 4 while a further 29.4 per cent of images passed criterion 5. 

\begin{figure}
	\includegraphics[width=\columnwidth]{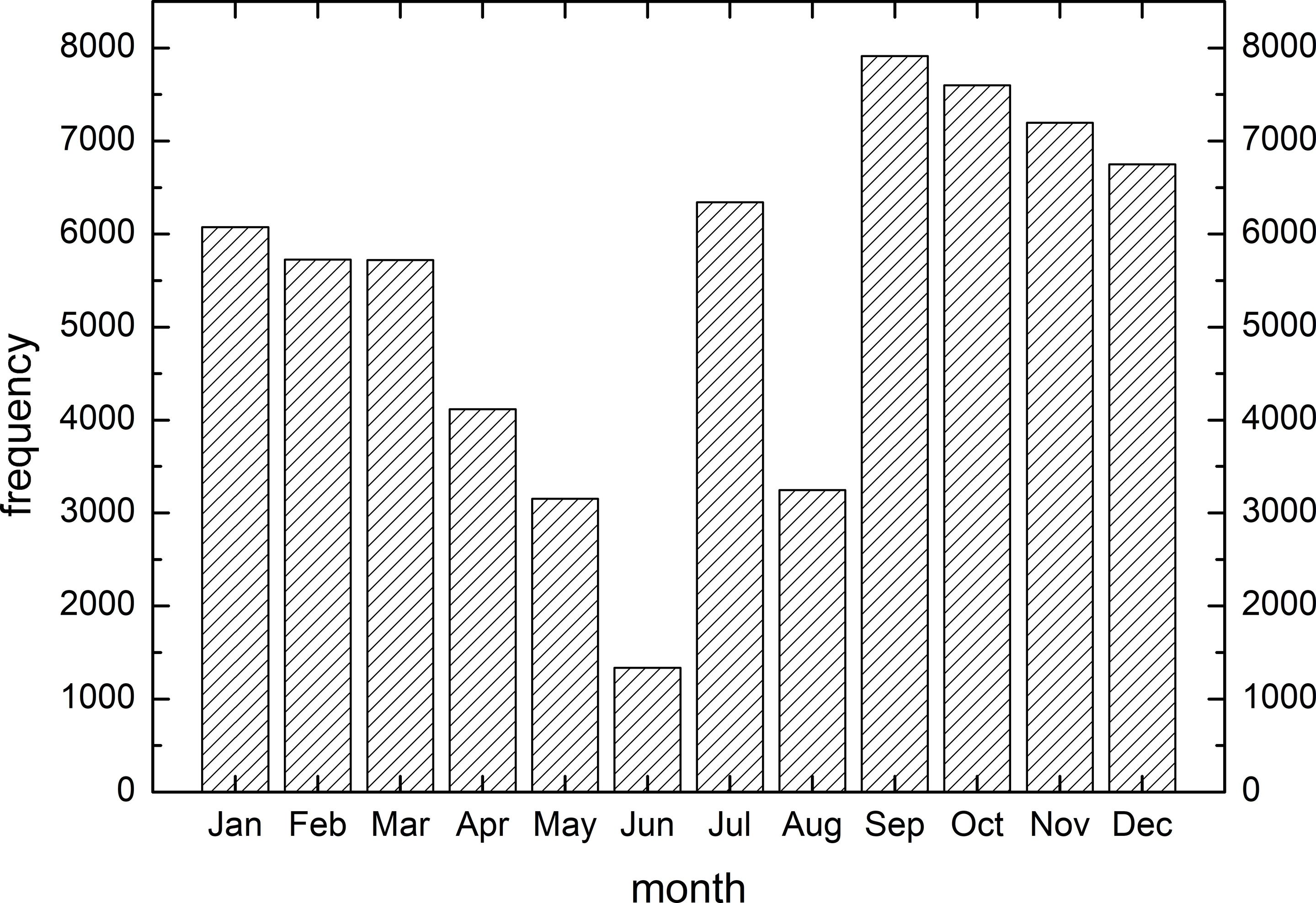}
    \caption{The distribution of data acquisition dates as a function of the month. This graph provides an overview of when the data was collected over the course of years.}
    \label{fig:statistics_month}
\end{figure}

\begin{figure}
	\includegraphics[width=\columnwidth]{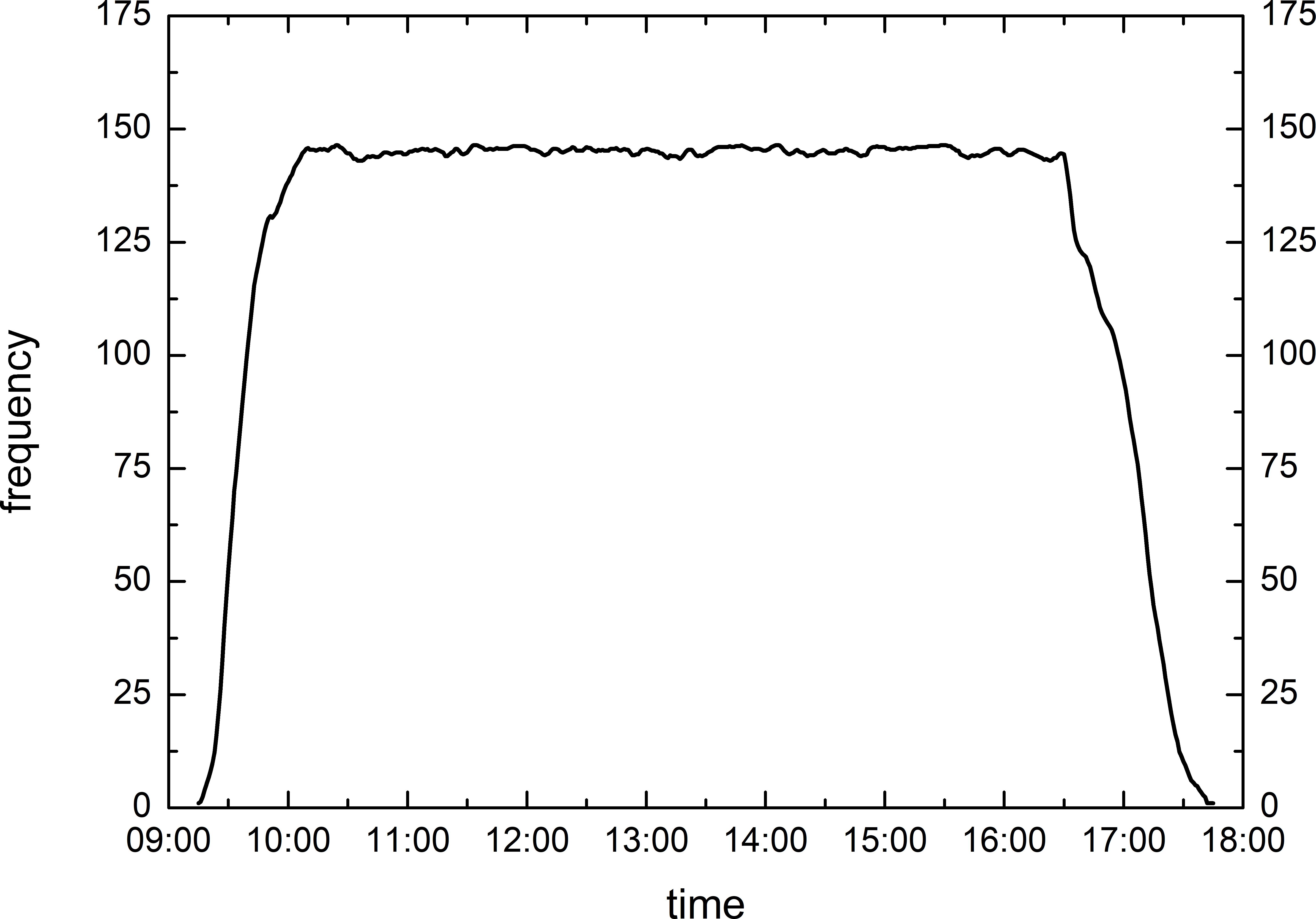}
    \caption{The distribution of data acquisition as a function of the time of day, binned to 5-minute intervals. This graph provides insights into the timing patterns of data collection throughout the day.}
    \label{fig:statistics_time}
\end{figure}

\begin{figure}
	\includegraphics[width=\columnwidth]{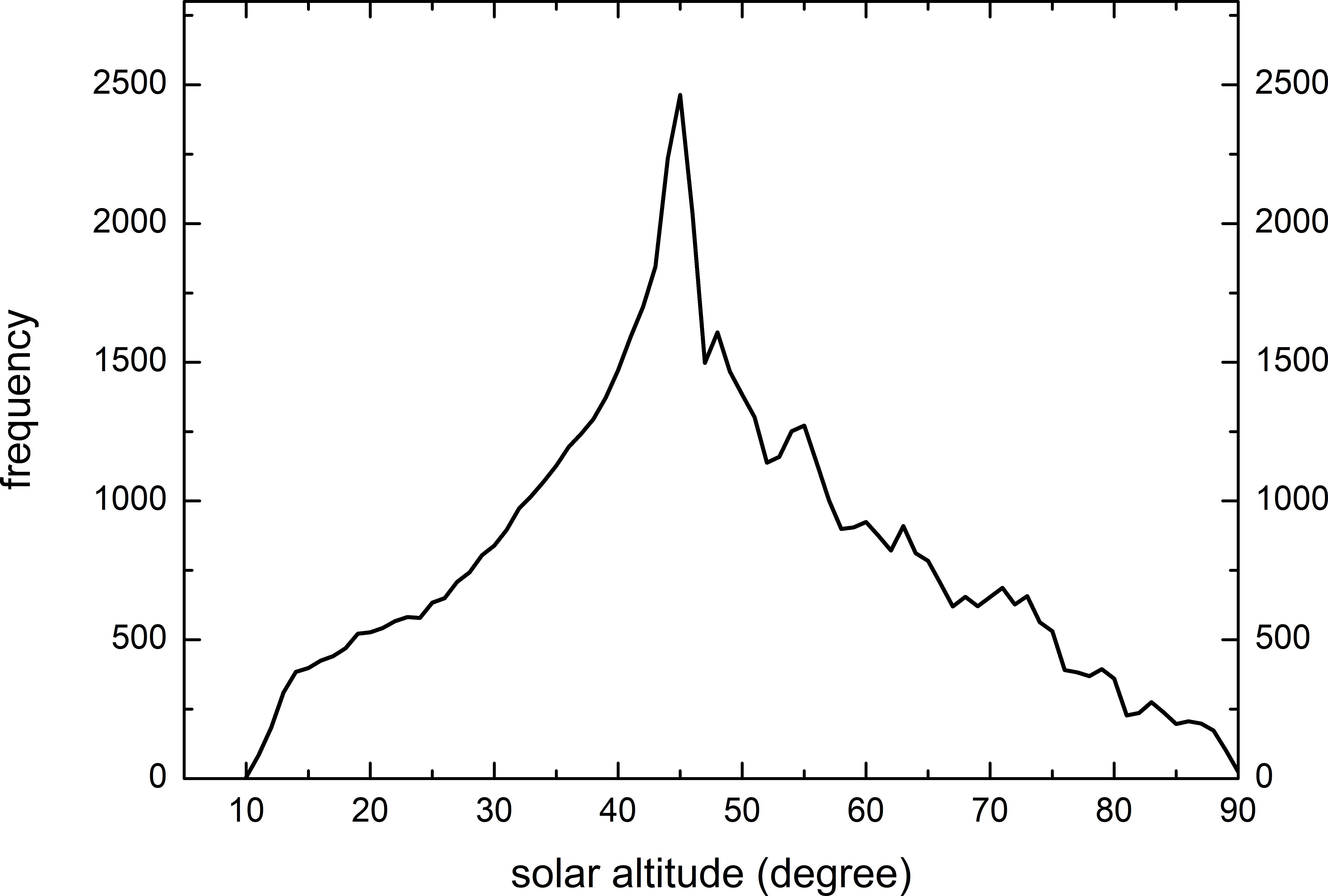}
    \caption{The distribution of data acquisition as a function of solar altitude. This graph provides an overview of the elevation angle of the Sun during the data collection period.}
    \label{fig:statistics_altitude}
\end{figure}

\begin{figure}
	\includegraphics[width=\columnwidth]{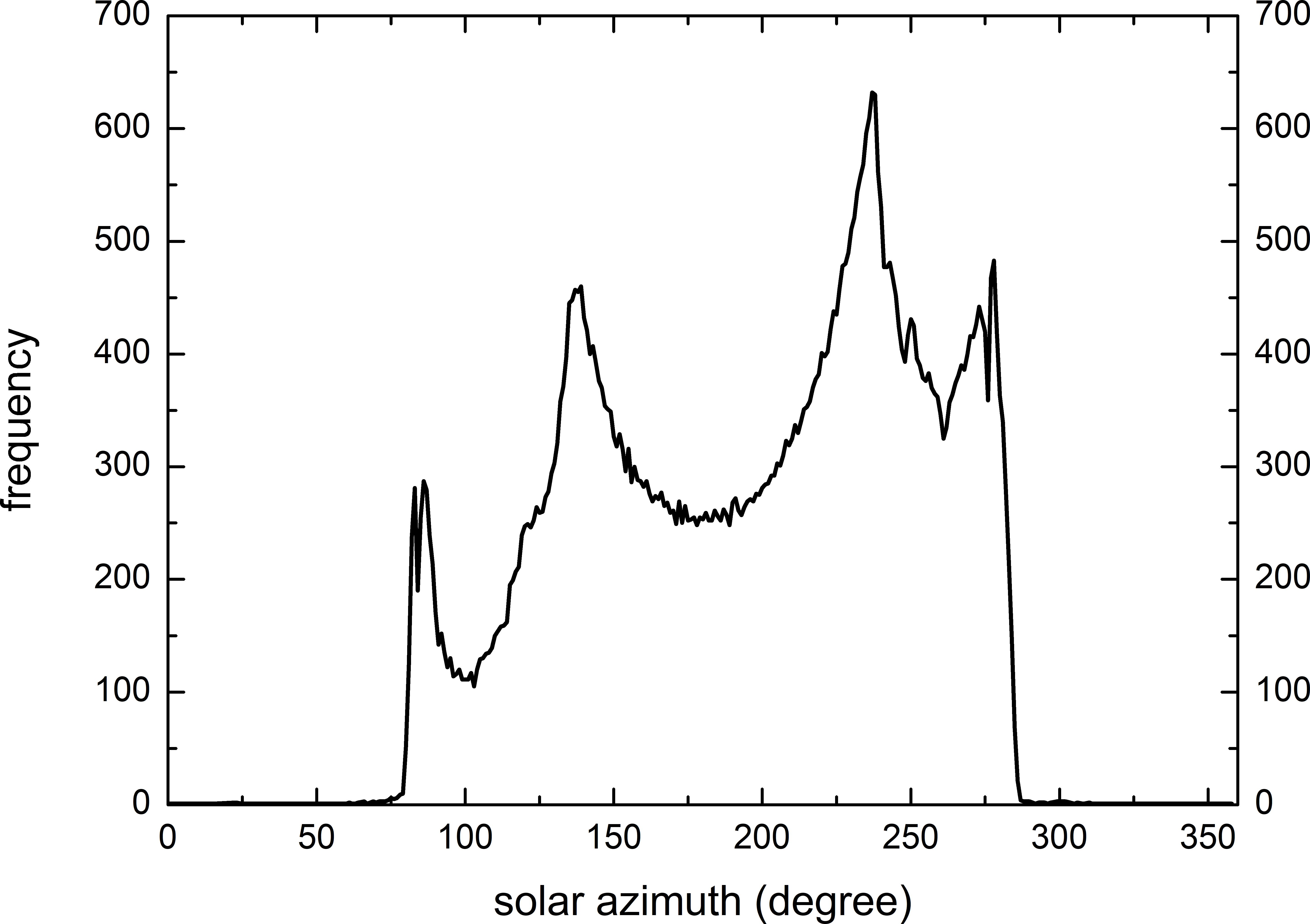}
    \caption{The distribution of data acquisition as a function of solar azimuth. This graph provides an overview of the  azimuthal position of the Sun during the data collection period.}
    \label{fig:statistics_azimuth}
\end{figure}

Based on the selection process, a total of 65,172 individual images were chosen for further analysis. These images represent approximately 22.6 per cent of the original dataset before applying any selection criteria. It is worth noting that these images cover a span of 168 individual days. To gain further insights into the selected images, statistical distributions were analysed against various parameters. Figures~\ref{fig:statistics_month}, ~\ref{fig:statistics_time}, ~\ref{fig:statistics_altitude} and ~\ref{fig:statistics_azimuth} present the distributions of the selected images in terms of month, time of day, solar altitude and azimuth, respectively. To compute the solar positions for the selected images, NASA Horizons System \citep{nasa:2021} was utilized. 

The statistical analysis reveals several interesting patterns in the data distribution:
\begin{itemize}
    \item Month histogram (Fig.~\ref{fig:statistics_month}): The sampling is biased towards autumn and winter (September to December), which are the dry seasons. This bias is likely due to favorable weather conditions during these months. Furthermore, the change in image acquisition interval from 1 minute to 5 minutes since 2023 January 19 has further reduced the sampling size before June.
    \item Time histogram (Fig.~\ref{fig:statistics_time}): The sampling is relatively even between 10:00 and 16:30 (local time, UTC+8 hr, hereafter). However, the sample size drops towards both ends of the day due to the filtering process based on selection criteria 1 to 3. Besides, the shorter daylight duration in winter contributes to the decrease in sample size after 16:30.
    \item Altitude histogram (Fig.~\ref{fig:statistics_altitude}): The histogram displays a sharp peak at 45 degrees, which corresponds to the maximum solar altitude during the winter season. 
    \item Azimuth histogram (Fig.~\ref{fig:statistics_azimuth}): The data is concentrated on the southern sky (azimuth range of $90^\circ$ to $270^\circ$), which is expected for a location in the subtropical region. The four peaks observed at around 9$0^\circ$, $135^\circ$, $240^\circ$ and $270^\circ$ can be further explained by referring to the polar plot in Fig.~\ref{fig:DVL_polar}.
\end{itemize}

\subsection{DVL of Solar Images}\label{sec:DVL}
Upon analyzing the VL values obtained from each solar image, it was observed that these values exhibit day-to-day variability. This variability can be attributed to fluctuations in atmospheric conditions or potential inconsistencies in the daily telescope's focusing position made by the operator. In order to accentuate the underlying trend of the VL values amidst these daily variations, a novel metric termed \textit{DVL} (\textit{D}ifference in \textit{VL}) was devised. DVL represents the disparity between the VL value of an individual image and the corresponding daily median VL value. By calculating DVL, we aimed to emphasize the relative changes in VL over time, thereby facilitating the identification of significant patterns or trends within the dataset.

\begin{figure*}
	\includegraphics[width=\columnwidth]{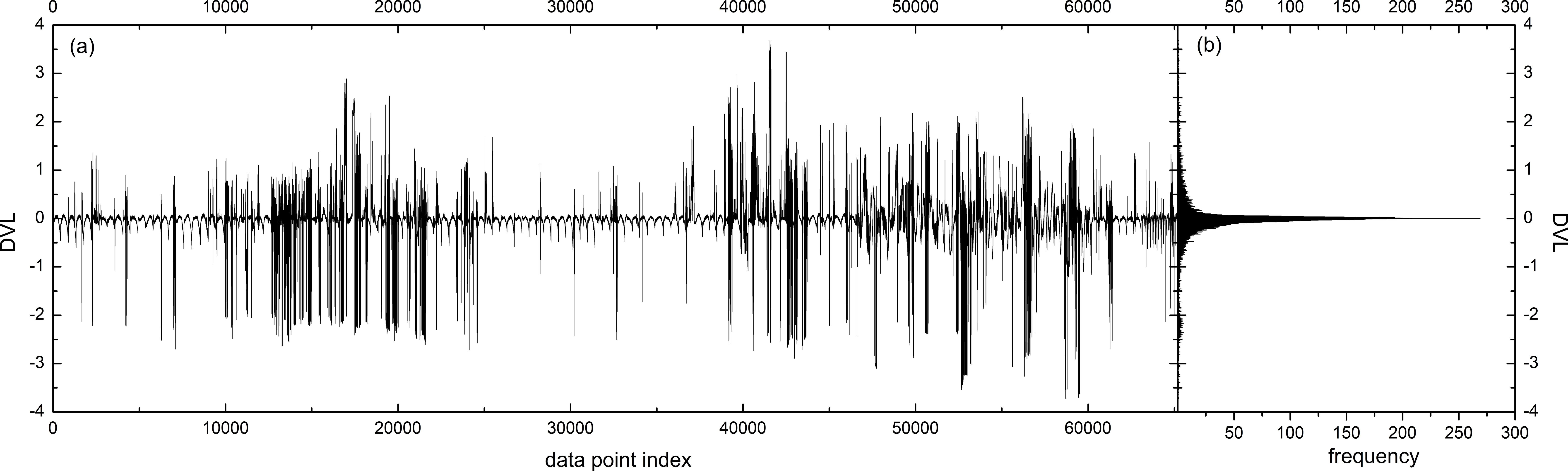}
    \caption{$(a)$ The time series and $(b)$ histogram of DVL}
    \label{fig:time_series}
\end{figure*}

Fig.~\ref{fig:time_series} showcases both the time series and histogram of DVL. Upon closer examination, the time series reveals a discernible daily rhythm that aligns with the movement of the Sun across the sky. Specifically, the DVL values are relatively small in the morning, reach their peak around noon when the Sun is at its highest point, and subsequently decrease towards the daily minimum near sunset. This pattern reflects the expected seeing variation throughout the day. Furthermore, the histogram of DVL demonstrates a prominent peak centered around $2.5\times10^{-4}$.

\begin{figure*}
	\includegraphics[width=\columnwidth]{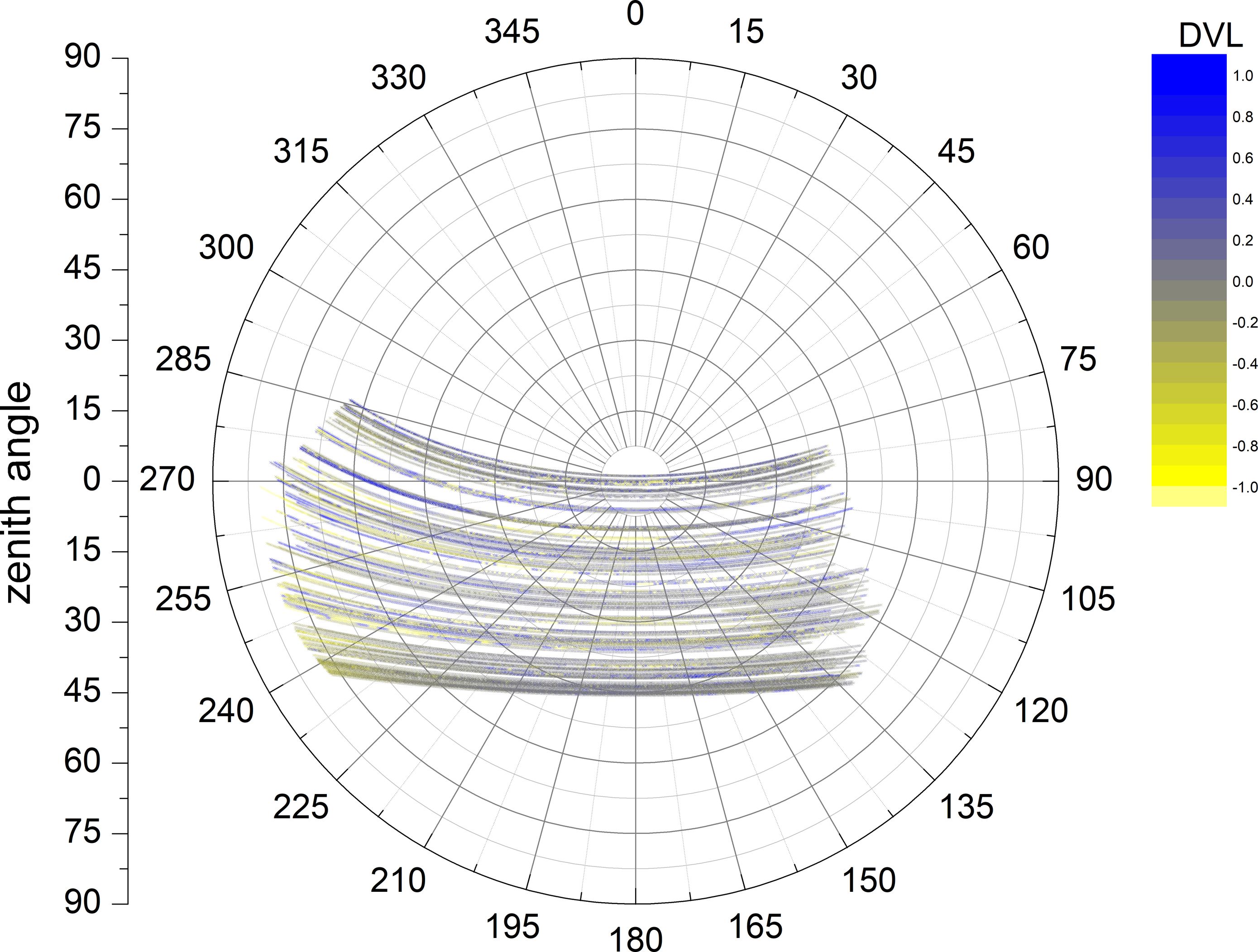}
    \caption{This polar plot of DVL provides a visualization of the data distribution and the corresponding DVL values across various solar positions, offering insights into the relationship between solar position and DVL magnitude.}
    \label{fig:DVL_polar}
\end{figure*}

Fig.~\ref{fig:DVL_polar} illustrates the distribution of DVL values in polar coordinates. In this representation, larger DVL values are depicted in a bluer shade, while smaller DVL values appear in a yellower shade. Upon examination, it becomes evident that the larger DVL values are primarily concentrated at higher altitudes. On the other hand, smaller DVL values are scattered across lower angles, indicating a correlation between DVL and solar altitude. 

The presence of peaks observed in Fig.~\ref{fig:statistics_azimuth} can now be reasonably explained. The peaks correspond to azimuth angles around 90$^\circ$, 135$^\circ$, 240$^\circ$ and 270$^\circ$, where the `line-of-sight' from the horizon to the zenith intersects a greater number of data points compared to other angles. This phenomenon aligns with the observation of increased data density along these particular azimuth directions.

No apparent azimuthal asymmetry is observed in the distribution of DVL, suggesting that there are no significant directional biases in the DVL values.

\subsection{DVL v.s. Solar Altitudes}\label{sec:altitude}
\begin{figure}
	\includegraphics[width=\columnwidth]{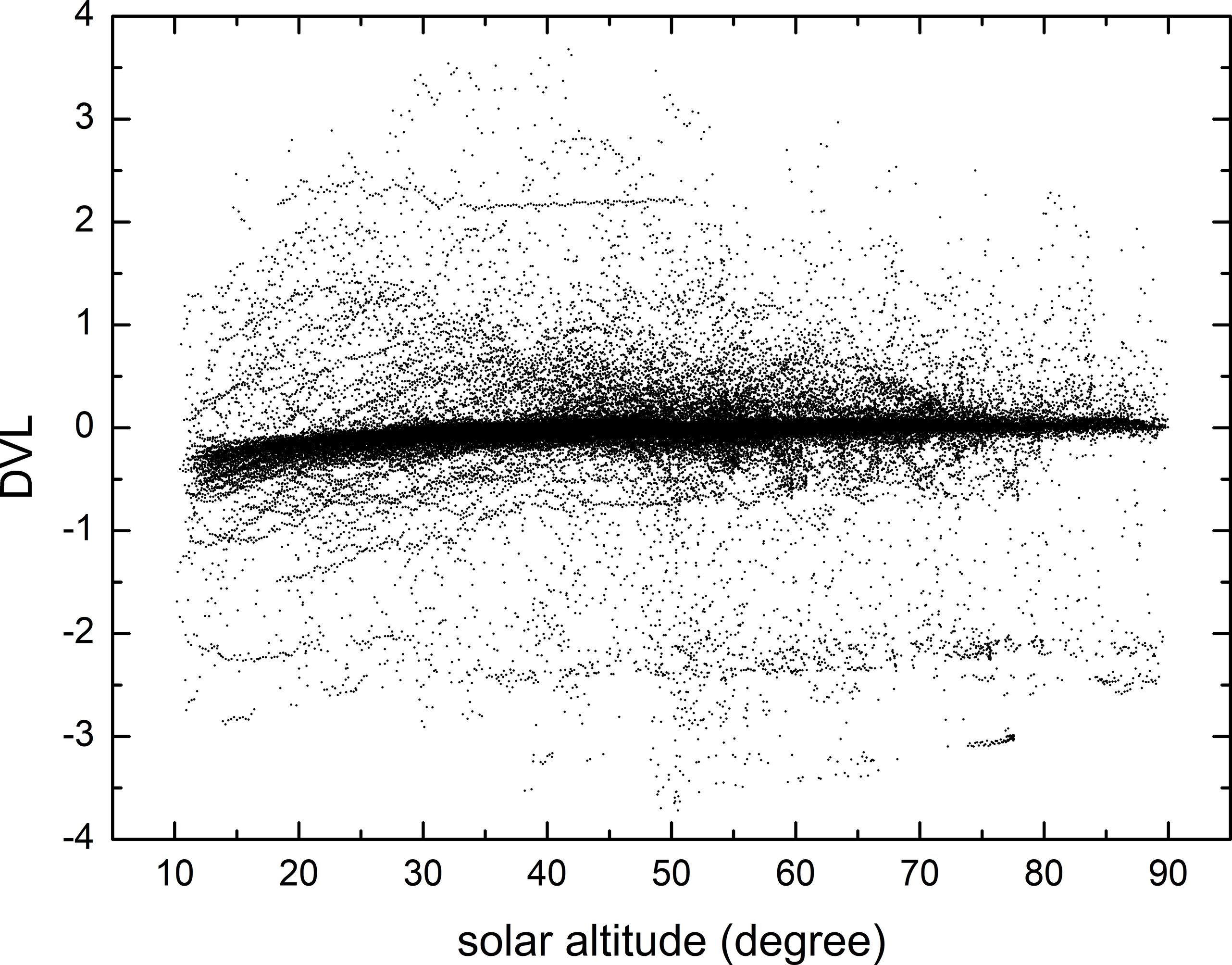}
    \caption{Scatter plot of DVL as a function of solar altitudes. The plot demonstrates how DVL values vary across different solar altitudes. There is a discernible trend of increasing DVL as solar altitudes rise, suggesting that solar images tend to exhibit greater sharpness when the Sun is positioned higher in the sky. Conversely, lower solar altitudes are associated with relatively lower DVL values, indicating reduced sharpness in the corresponding images.}
    \label{fig:all_altitude}
\end{figure}

Fig.~\ref{fig:all_altitude} displays a scatter plot depicting the relationship between DVL and solar altitudes. The plot reveals a weak trend of DVL increasing as solar altitude rises. This observation suggests that solar images tend to exhibit greater sharpness when the Sun is positioned higher in the sky, and conversely, lower solar altitudes are associated with relatively reduced sharpness. 

Beyond approximately $40^\circ$ solar altitude, the central data distribution reaches a plateau, indicating that the variation in DVL becomes relatively stable. This trend suggests that the image quality does not significantly degrade when the solar altitude exceeds this angle in general. In other words, approximately 65 per cent of the observation time is expected to yield solar images of satisfactory quality, disregarding the potential impact of cloud coverage. These findings imply that during a substantial portion of the observation period, the quality of the captured solar images remains reasonably acceptable, contributing to sound data acquisition for analysis and research purposes.

\subsection{Image Sharpness Function, ISF}\label{sec:isf}
Image degradation is attributed to starlight's atmospheric attenuation. This degradation is typically quantified using airmass ($X$), which is formulated based on the thickness of the air column along the line of sight to the target. Specifically, we express airmass as a function of the zenith angle ($z$) of the target, where $z$ is defined as $90^\circ$ minus the altitude. 

In this section, our objective is to develop an Image Sharpness Function (ISF) based on our observational data. ISF serves as a means to characterize the relationship between image sharpness and airmass, providing a quantitative measure of the degradation in image quality due to atmospheric attenuation. To accomplish this, we referred to existing airmass formulae available in the literature as a reference point.

\begin{figure}
	\includegraphics[width=\columnwidth]{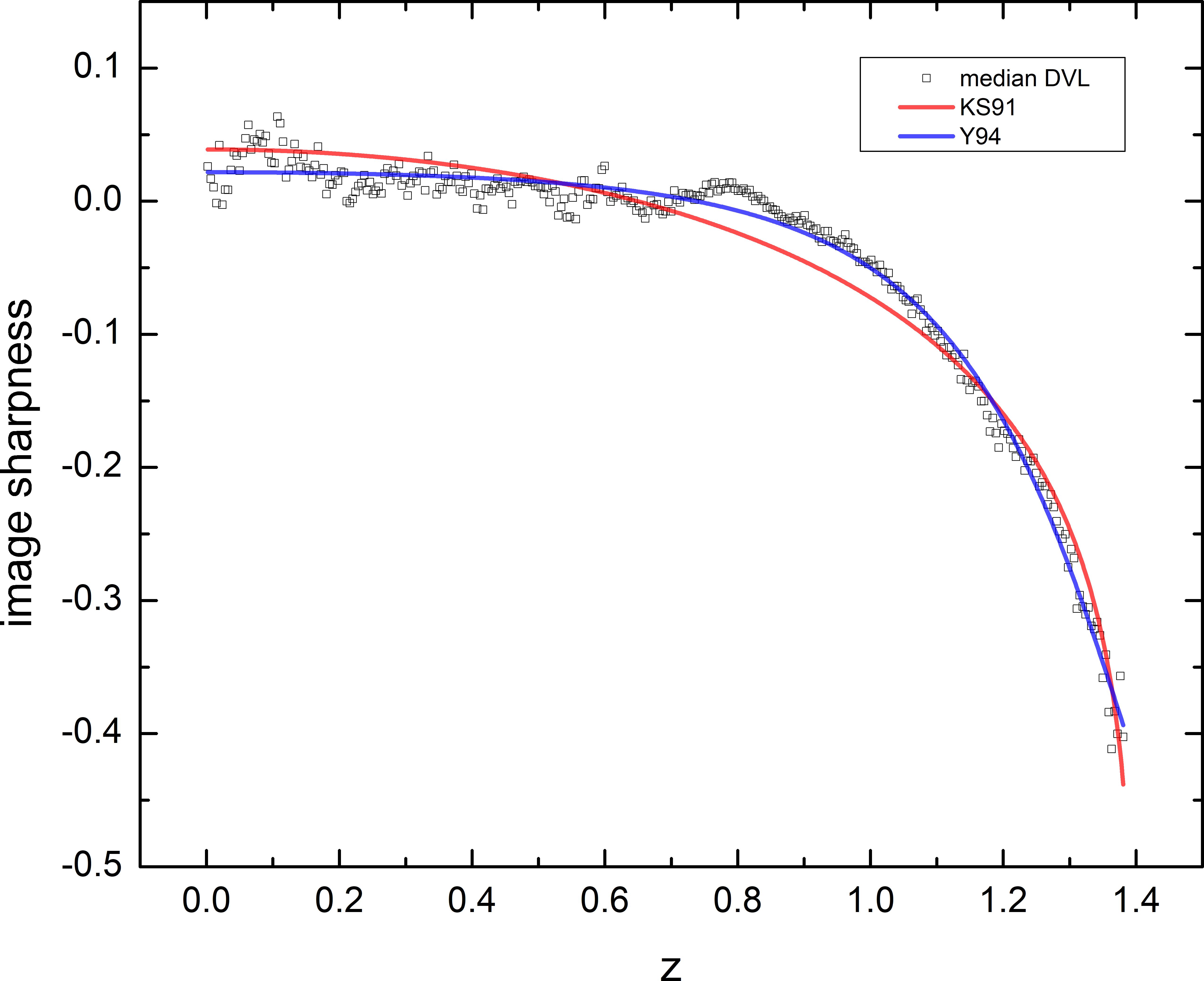}
    \caption{The trend of image sharpness is evident from the median DVL observations (depicted as \textit{black squares}) at various zenith angles $z$. The red curve, derived from the airmass formula in \citetalias{ks91}, exhibits a reasonable fit to the data points. Similarly, the blue curve, which corresponds to the airmass formula in \citetalias{young:1994}, also demonstrates a good alignment with the observed median DVL values.} 
    \label{fig:DVL_z}
\end{figure}

Existing formulae expresses $z$ in radian, therefore we initially converted the solar altitudes (in degrees) in our dataset to $z$ values (in radians). Subsequently, we plotted the median DVL for each $0.25^\circ z$ interval, as depicted in Fig.~\ref{fig:DVL_z}. It is important to note that the last three data points with $z>1.381$ (corresponding to solar altitudes less than $10.88^\circ$) were excluded from the analysis due to under-sampling. Each of these bins contained fewer than 10 samples, rendering them insufficient for meaningful statistical analysis.
 
We evaluated two distinct forms of ISF. The first form is as follows:
\begin{equation}
    DVL = a - [(1- b\sin^2(z)]^c 	
    \label{eq:KS91airmass}
\end{equation}
which aligns with the airmass formula proposed by \citet{ks91} (referred to as \citetalias{ks91} hereafter): $X = [1-0.96\sin^2(z)]^{-0.5}$.

By applying the Levenberg-Marquardt method \citep{press:2007}, we fitted the median DVL observations to Eq.~\ref{eq:KS91airmass}, resulting in a well-fitted curve with a high coefficient of determination ($R^2$=0.962).

We also explored a second form of ISF, which can be represented as:
\begin{equation}
    DVL = \frac{d\cos(z) + e}{f\cos^2(z) + g\cos(z) + h} 	
    \label{eq:young94airmass}
\end{equation}
according to the low accuracy airmass formula in \citet{young:1994} (referred to as \citetalias{young:1994} hereafter) which utilizes parameter values of 1.003198, 0.101632, 1, 0.090560 and 0.003198 for $d$, $e$, $f$, $g$ and $h$ respectively. We repeated the fitting process using Eq.~\ref{eq:young94airmass} and obtained a curve with a slightly improved coefficient of determination ($R^2$=0.985). This indicates a stronger correlation between the predicted sharpness values based on this form of ISF and the actual median DVL observations.

\begin{table}
	\centering
	\caption{Best-fitted values of the parameters in two different forms of ISF}
	\label{tab:best_fit}
	\begin{tabular}{ccc} 
		\hline
		ISF form & Parameter & Best-fitted value \\ 
		\hline
                                       & $a$  &  1.039 \\
        $a - [(1- b\sin^2(z)]^c$       & $b$  &  1.029 \\ 
                                       & $c$  & -0.081 \\
                                       \hline
                                                                & $d$      & -3.256  \\
                                                                & $e$      & 2.412   \\
        $\frac{d\cos(z) + e}{f\cos^2(z) + g\cos(z) + h}$ & $f$             & -40.035 \\ 
                                                                & $g$      & 5.209   \\ 
                                                                & $h$      & -4.123  \\
		\hline
	\end{tabular}
\end{table}

Table~\ref{tab:best_fit} provides the fitted parameters obtained from the regression analysis. Fig.~\ref{fig:DVL_z} illustrates the best-fitted curves derived from the ISF. Taken as a whole, it becomes apparent that the overall trend of DVL aligns more closely with the variation in airmass as described by the \citetalias{young:1994} formula. However, it is worth noting that for larger zenith angles ($z$) beyond 1.36 (equivalent to $78^\circ$), the form of the \citetalias{ks91} airmass formula fits better. This observation is consistent with the findings highlighted by \citetalias{young:1994}, who noted that accurately formulating airmass near the horizon presents notable challenges due to its sensitivity to various atmospheric conditions, such as surface temperature, pressure and the vertical structure of the atmosphere.

\section{Conclusions \& Discussion}
\label{sec:discussion}
Our study has demonstrated the feasibility, efficiency and simplicity of using the Variance of Laplacian (VL) method to analyse the quality of solar images. This method assigns a `score' to each image based on the sharpness of the solar disk, providing a quantitative measure of image quality and, consequently, atmospheric conditions, such as seeing.

While the measurement of seeing is typically achieved by estimating the angular size of the point spread function of stellar images or using specialized equipment like seeing monitors, our method utilizes standard solar images. Our method can be employed to extract the diameter of the solar disk from individual images. This approach enables a direct and expeditious estimation of the site's seeing parameter $r_0$ by analyzing variations in the measured solar diameters, as illustrated in a recent paper by \citep{duan:2024}. It eliminates the need for time-consuming data reduction processes, extensive computational resources, or additional investments in instrumentation. In the future, we plan to expand our method by testing it with disk images of the Moon, bright planets and defocused stars. This will allow us to develop real-time image quality monitors or seeing monitors that can be executed on-the-spot.

Our analysis has reproduced a well-known trend indicating a relationship between altitude and image quality: higher targets in the sky (with lower airmass) exhibit better image quality, while lower targets have poorer image quality. In particular, the quality of solar images taken above $40^\circ$ altitude remains reasonably acceptable in general. We have represented the relationship through an Image Sharpness Function (ISF), which describes the image quality as a function of the target's zenith angle. Although we have only tested two common forms of airmass formulas, our analysis strongly suggests that ISF closely resembles the behavior of airmass, indicating a deep relationship between the two. By studying image quality, we hope to facilitate the challenging task of modeling airmass.

Each observing site possesses unique characteristics, including its own ISF. Developing a site-specific ISF offers at least two benefits. Firstly, it can be incorporated into calculation tools like the SKYCALC Sky Model Calculator\footnote{\url{https://www.eso.org/observing/etc/bin/gen/form?INS.MODE=swspectr+INS.NAME=SKYCALC}} to enhance airmass or seeing estimations for observers' reference during the preparation of observing runs. Secondly, by regularly monitoring ISF rather than relying on separate seeing monitors, one can continuously assess the observing site conditions over an extended period. Furthermore, if archival images are available, retrospective analysis becomes possible.  

Future research should focus on enhancing the accuracy and robustness of the VL algorithm. This can be achieved by testing alternative edge detection algorithms \citep{song:2018}, noise filtering algorithms \citep{wang:2007, jain:2016} and employing machine learning techniques.

Lastly, this study highlights the value of the vast collection of archival solar images, which can serve as a valuable data repository for research beyond solar physics. 

\section*{Data Availability}
Data and codes available on request.

\section*{Declaration of Competing Interest}
None.

\section*{Acknowledgements}
E. Yuen acknowledges the support received from the Department of Physics at The University of Hong Kong. We would also like to express our gratitude to the reviewers for their valuable time and effort in reviewing the manuscript. Their feedback and suggestions have significantly contributed to the improvement of this study.

\bibliographystyle{cas-model2-names}
\bibliography{solar3} 

\end{document}